\documentclass{iopart}

\usepackage[english]{babel}
\usepackage{verbatim}
\usepackage{graphicx}

\begin{document}
\title[Ground state capture in $^{14}$N(p,$\gamma$)$^{15}$O at LUNA]{Ground state capture in $^{14}$N(p,$\gamma$)$^{15}$O studied above the 259 keV resonance at LUNA}

\author{H P Trautvetter$^1$, D Bemmerer$^3$, R Bonetti$^6$, C Broggini$^8$, A Caciolli$^8$, F Confortola$^4$, P Corvisiero$^4$, H Costantini$^4$, \mbox{Z Elekes$^2$}, A Formicola$^5$, Zs F$\ddot{\mathrm{u}}$l$\ddot{\mathrm{o}}$p$^2$, G Gervino$^{10}$, \mbox{A Guglielmetti$^6$},
Gy Gy$\ddot{\mathrm{u}}$rky$^2$, C Gustavino$^5$, G Imbriani$^7$, \mbox{M
Junker$^5$}, A Lemut$^4$, B Limata$^7$, M Marta$^3$, \mbox{C Mazzocchi$^6$,} R
Menegazzo$^8$, P Prati$^4$, V Roca$^7$, C Rolfs$^1$, C Rossi Alvarez$^8$, E
Somorjai$^2$, O Straniero$^9$, F Strieder$^1$, \mbox{F Terrasi$^7$,} S
Vezz\`u$^{11}$ and A Vomiero$^{12}$}

\address{$^1$ Institut f$\ddot{\mathrm{u}}$r Experimentalphysik III, Ruhr-Universit$\ddot{\mathrm{a}}$t Bochum, Germany}
\address{$^2$ Institute of Nuclear Research (ATOMKI), Debrecen, Hungary}
\address{$^3$ Forschungszentrum Dresden-Rossendorf, Dresden, Germany}
\address{$^4$ Universit\`a di Genova and INFN Sezione di Genova, Genova, Italy}
\address{$^5$ INFN, Laboratori Nazionali del Gran Sasso (LNGS), Assergi (AQ), Italy}
\address{$^6$ Istituto di Fisica Generale Applicata, Universit\`a di Milano and INFN Sezione di Milano, Italy}
\address{$^7$ Dipartimento di Scienze Fisiche, Universit\`a di Napoli "Federico II" and INFN Sezione di Napoli, Napoli, Italy}
\address{$^8$ Istituto Nazionale di Fisica Nucleare (INFN), Sezione di Padova, Italy}
\address{$^9$ Osservatorio Astronomico di Collurania, Teramo, and INFN Sezione di Napoli, Napoli, Italy}
\address{$^{10}$ Dipartimento di Fisica Sperimentale, Universit\`a di Torino and INFN Sezione di Torino, Torino, Italy}
\address{$^{11}$ CIVEN Nano Fabrication Facility, Venezia, Italy}
\address{$^{12}$ INFM-CNR SENSOR Lab, Brescia and INFN Laboratori Nazionali di Legnaro, Italy}
\address{\bf (The LUNA collaboration)}


\begin{abstract}
We report on a new measurement of $^{14}$N(p,$\gamma$)$^{15}$O for the ground
state capture transition at $E_p$ = 360, 380 and 400 keV, using the 400 kV LUNA
accelerator. The true coincidence summing effect --the major source of error in
the ground state capture determination-- has been significantly reduced by using a
Clover--type gamma detector.
\end{abstract}

\section{Introduction}
\begin{figure}[tbp]
    \centering
        \includegraphics[angle=-90,width=0.49\textwidth]{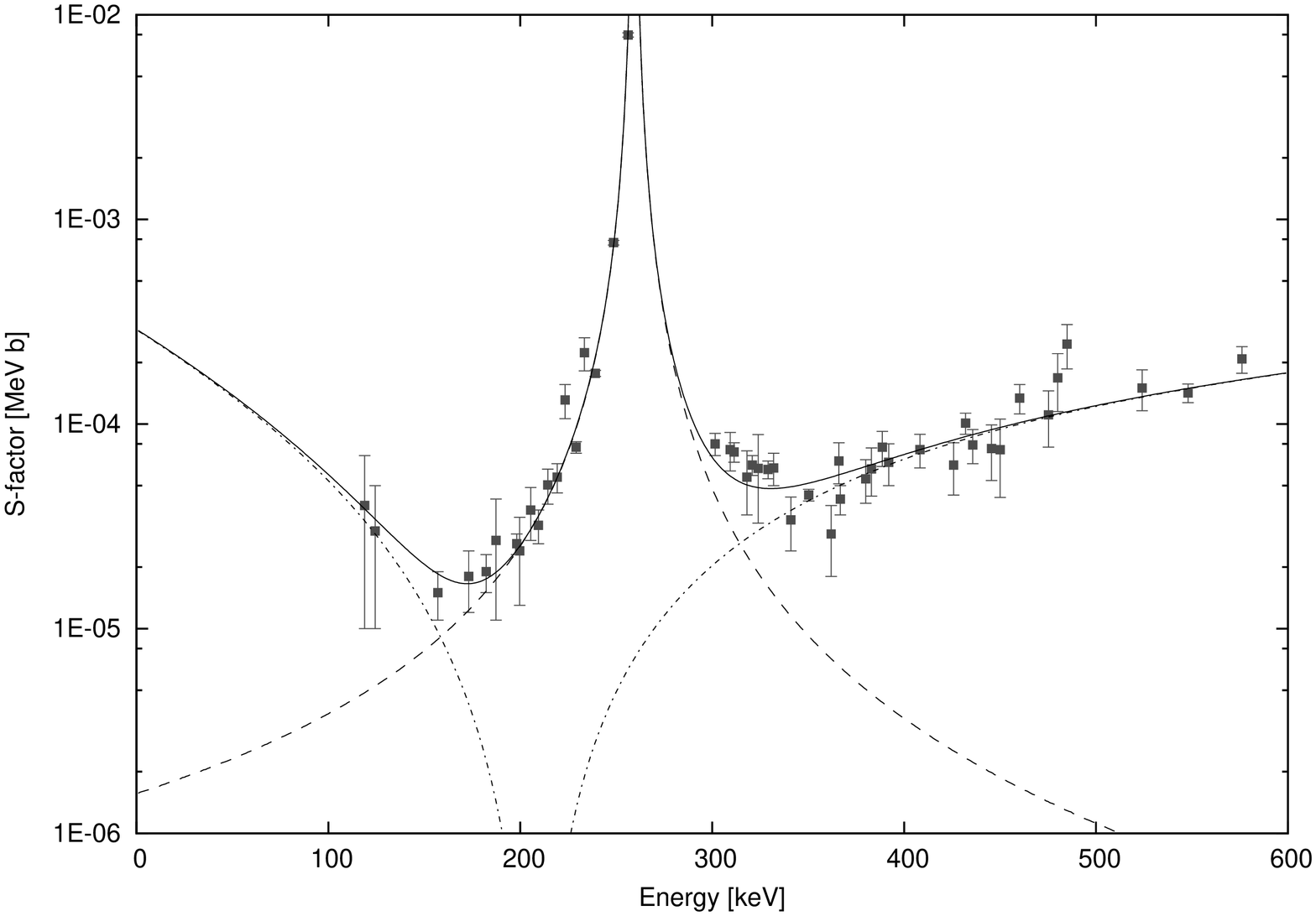}
        \includegraphics[angle=-90,width=0.49\textwidth]{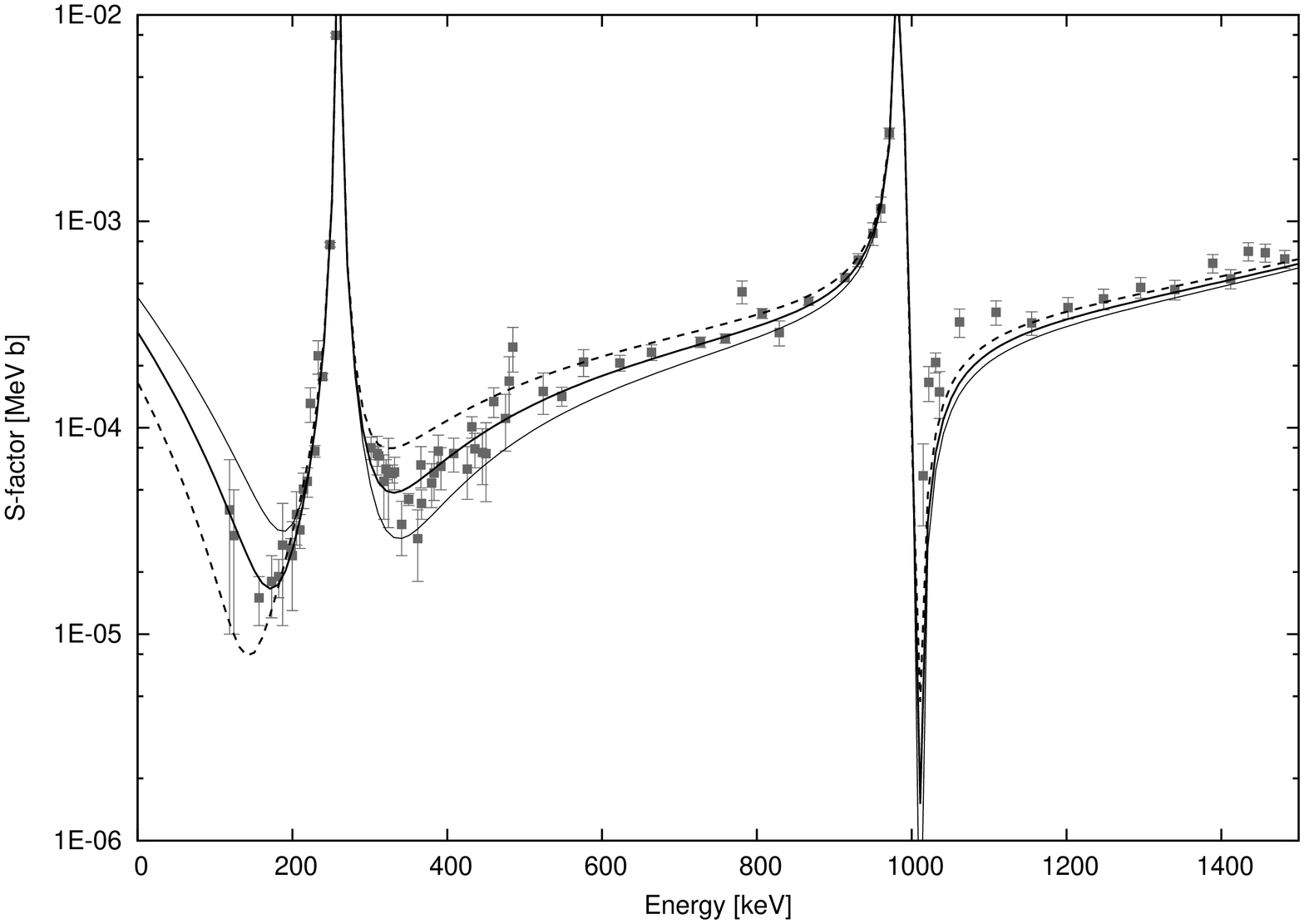}
    \caption{Astrophysical S--factor for radiative proton capture to the ground state of $^{15}$O as a function of center of mass energy. The left panel shows data from \cite{form+imbr,runkle,schroeder} together with a previous R--matrix fit \cite{form+imbr}. Shown is also the contribution from the 259 keV resonance alone (dashed line) and the contribution primarily from the subthreshold state and the external contribution (dot--dashed line). The right panel illustrates the influence of change in $\Gamma_{\gamma}$ of the subthreshold state where $\Gamma_{\gamma}$ ranges from 0.6 (dashed line), 0.8 (solid line) to 1 eV (solid thin line).}
    \label{fig:1}
\end{figure}

The $^{14}$N(p,$\gamma$)$^{15}$O reaction ($Q$ = 7297 keV) is the slowest process
in the hydrogen burning CNO cycle \cite{rolfs} and thus of high astrophysical
interest. This reaction plays a role for the neutrino spectrum of the Sun
\cite{bahcall} as well as in the age determination of globular clusters
\cite{glob_cluster}. The reaction was recently studied in three experiments at
energies ranging from $E_{cm}$ = 70 to 480 keV [4--6] and before over a wide range
of energies, i.e. 240 to 3300 keV (\cite{schroeder} and references therein). A
significant reduction of the ground state contribution \cite{angulo} has been
found \cite{form+imbr,runkle}. However, the analysis was hampered by the fact that
the usage of large detectors in close geometry has as a consequence that the
"ground state contribution" is masked by summing--in due to coincidence events
from the cascade transitions in $^{15}$O. Necessary corrections were of the order
of a factor two to four, thereby increasing the uncertainty
\cite{form+imbr,runkle}. Moreover, an R--matrix analysis revealed (figure
\ref{fig:1}, left panel) that below the 259 keV resonance the data followed
primarily the low energy wing of the resonance. These data could not probe the
behavior of the interference structure, which is needed for reliable
extrapolation, due to a minimum of the S--factor curve near $E_{cm}$ = 160 keV.

The total S--factor at 70 keV is known with $S$(70) = 1.74$\pm$0.14 keV b
statistical error \cite{lemut}. However, the ground state contribution at that
energy is expected to be \cite{form+imbr} $S_{gs}$(70) = 0.07 keV b, so the
summing crystal study in Ref. \cite{lemut} is not sufficiently sensitive to probe
the contribution of the ground state transition. The value of $S_{gs}$(0) depends
on the $\Gamma_{\gamma}$ of the subthreshold state at 6.79 MeV excitation energy
in $^{15}$O. The influence of a change in $\Gamma_{\gamma}$ of the subthreshold
state is illustrated in the right panel of figure \ref{fig:1} where
$\Gamma_{\gamma}$ ranges from 0.6 (dashed line), 0.8 (solid line) to 1 eV (solid
thin line). When lowering the width of the subthreshold state $S_{gs}$(0)
decreases, the destructive interference minimum moves to lower energies and hence
recovers much earlier at higher energies, i.e. the cross section is expected to be
larger at energies above the 259 keV resonance. For the above range in
$\Gamma_{\gamma}$ one expects a change in cross section of a factor of three
around 330 keV. Also from figure 1 it is obvious that the non--resonant shape
could be studied again above $E_{cm}$ = 300 keV. We have therefore designed an
experiment in the energy range 300 to 400 keV using a BGO shielded Clover detector
\cite{clover} to reduce significantly the summing--in contributions from the true
coincidence events from the $^{15}$O cascade transitions.

\section{Experiment}
The set up was similar to Ref. \cite{form+imbr} with the detector placed at
55$^{\circ}$ with respect to the beam axis. The 400 kV LUNA 2 accelerator
\cite{formicola} provided proton beam currents of up to 350 $\mu A$ on target. The
N--targets were produced by reactive plasma deposition of TiN onto Ta backings
with observed energy loss of 50 keV for protons at $E_p$ = 280 keV in the TiN
layer. A BGO shielded Clover detector \cite{clover} with the crystals front faces
at a distance of 9 cm from the target was used in the present experiment in a
simplified way. The individual events of the four segments of the clover were
collected in single ADC's. In addition, the analogue signals of the individual
segments were summed (hard sum) and stored in another independent ADC which was in
anti--coincidence with the surrounding BGO crystals for normal running conditions.
While the sum of the individual segments after calibration and gain matching (soft
sum) should provide the greatest reduction in the summing effect, the hard sum
acts like a large crystal (at 9 cm distance as opposed to 1.5 cm \cite{form+imbr}
and 0.9 cm \cite{runkle}) with Compton background reduction due to the BGO
anti--coincidence circuit and reduction in the summing by the larger distance.
Measurements with radioactive sources ($^{137}$Cs, $^{60}$Co) placed at the target
position, have been performed to gain absolute efficiency information. Losses due
to the cascade structure of the gamma events through the anti--coincidence circuit
with the surrounding BGO's have been studied in the same manner, showing less than
5\% effect. This has been corrected for in the data analysis.

\subsection{Target profile}
\label{sec:target} The target profile was expected to deteriorate after heavy
bombardment with protons. Therefore the profile was checked every day (after about
25 C proton irradiation) by a scan of the 278 keV resonance. There has been no
significant change of the observed resonance energy, hence no relevant C--build up
was noted during the whole course of the experiment due to an LN$_2$ cooled shroud
placed before the target. One can, however, observe the following important
parameter changes: i) the thickness of the target reduces with time; ii) the rear
tail width increases; and iii) the integral over the target decreases. The
thickness, tail-width and integral over the profile behaved nearly linearly with
the total accumulated dose on the target and have been corrected for in the
analysis. After at most 40\% reduction in the thickness, the target was replaced.

\subsection{Efficiency determination and summing corrections}
\begin{figure}[tbp]
    \centering
        \includegraphics[angle=-90,width=1.0\textwidth]{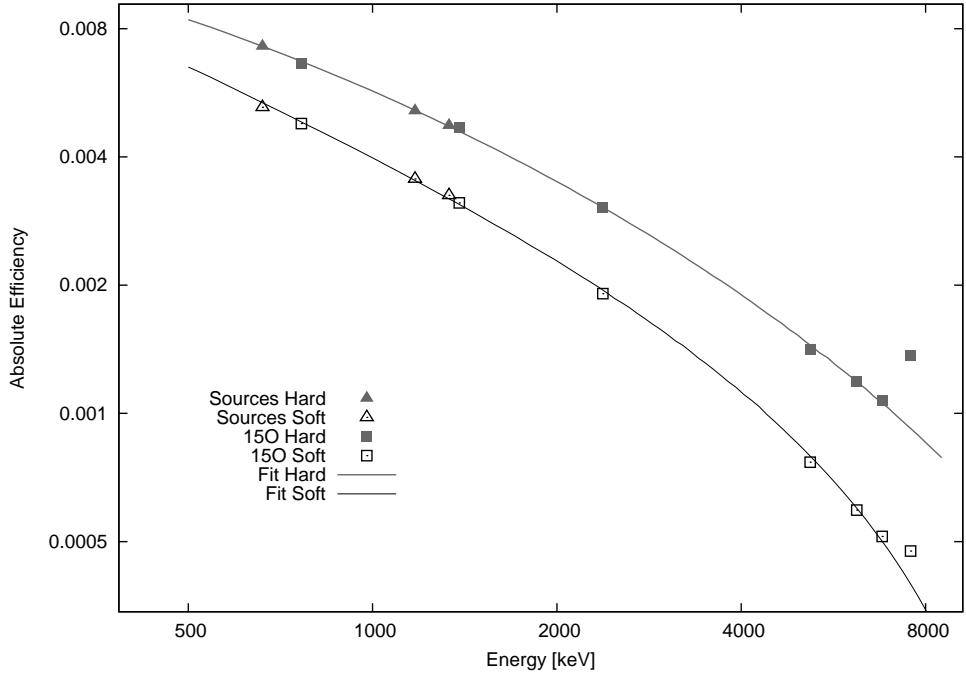}
    \caption{Absolute efficiency curves for both hard sum (upper points) and soft sum (lower points): please note the difference caused by the summing effect at E$_{\gamma}$ = 7.6 MeV.}
    \label{fig:2}
\end{figure}

Absolute and relative efficiency of the Clover detector were determined by source
measurements and the $\gamma$--rays coming from the 259 keV resonance in
$^{14}$N(p,$\gamma$)$^{15}$O, respectively. Here, the branching ratios and $\omega
\gamma$ value of the 259 keV resonance were used \cite{form+imbr,lemut}. A free
parameter was the stoichiometry $y/x$ of the Ti$_x$N$_y$ target. The ratios of the
primary and secondary transitions were normalized to the source results and thus
extending the range of energies from 662 to 6791 keV. As expected, the ground
state transition showed sizable summing contributions in the hard sum spectra
(figure \ref{fig:2}). The effect is, however, much lower than the factor 3.5
\cite{form+imbr} or even higher \cite{runkle} observed in previous work at low
energy. This effect was again reduced in the soft sum efficiency (fig.
\ref{fig:2}). The summing correction for the energy range of the present
experiment (i.e. well above the resonance) is expected to be less then 20\% for
the hard sum and less then 5\% for the soft sum, taking into account the previous
cross section results for the ground state transition and the transitions through
6.18 and 6.79 MeV state. Hence, even an uncertainty of 10\% in that correction
would lead to less than 2\% error in the final result. Finally, the summing--out
effects for the cascade transitions (e.g. the 6.79 MeV $\gamma$--ray) can be
estimated to be also less than 2\% for the hard sum.

\subsection{Relative determination of the ground state S--factor}
\label{sec:relative} The aim of the present experiment was to determine the cross
section at energies above the 259 keV resonance (fig. \ref{fig:1}). We determined
the cross section relative to the well studied transition to the 6.79 MeV state.
Such a procedure has several advantages:
\begin{itemize}
    \item The measurement is independent of the knowledge of absolute quantities such as
target stoichiometry, target profile, charge, stopping power and absolute
efficiency.
    \item The mean value of cross section for capture to the 6.79 MeV state of recent publications \cite{form+imbr,runkle} agrees within 2\%, hence no large
uncertainty arise from such normalisation.
    \item The effective energy determination is not critical due to similar
energy dependence of both cross sections which are controlled by the
penetrability.
    \item The relative efficiency determination is not affected by large systematic
errors when studying the secondary transition, the 6.79 MeV $\gamma$--line which
is near the expected ground state transition.
\end{itemize}
The peak contents were obtained by subtracting a fitted linear background in the
region surrounding the peak of interest. Effective energies were obtained by
determining the centroid of the observed non resonant ground state transition. The
resonance contribution through the tail of the target profile can be subtracted by
extracting the resonance part from the 765 keV primary line and correcting for the
respective efficiency. A cross check of the validity of this procedure is the
comparison of the efficiency corrected yield in the 6.79 MeV peak with the sum of
the 776 keV resonant and non resonant part (shifting with proton energy). Good
agreement was found after subtraction of background lines in the broad non
resonant part of the primary line. This indicates that no problem arises due to
incorrect angular position for the detector set up with respect to the beam axis.

\begin{figure}[tbp]
    \centering
        \includegraphics[angle=-90,width=1.0\textwidth]{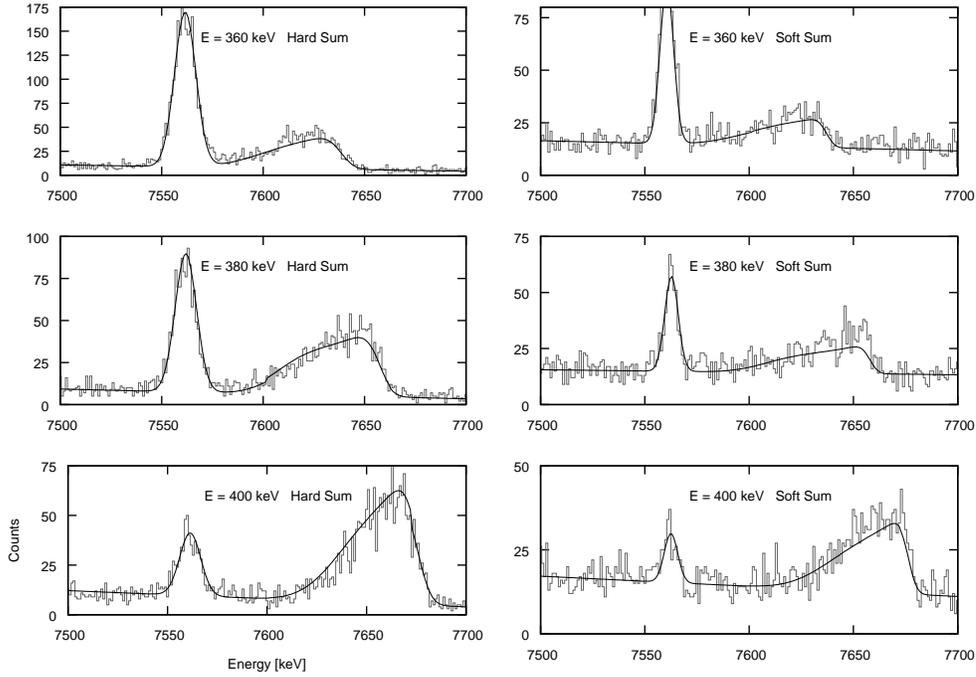}
    \caption{Line shape analysis result for all energies; left is shown the hard sum results and right the soft sum results.}
    \label{fig:3}
\end{figure}

\subsection{Line shape analysis}
\label{sec:LineShape} In addition, the cross section can be determined
independently through a line shape analysis (fig. \ref{fig:3}). The shape of the
non resonant reaction $\gamma$--lines contains information on the cross section
behavior over the target thickness. As an approximation, second order polynomial
function of the previous R--matrix analysis \cite{form+imbr} was sufficient to
describe the energy dependence of the cross section in the energy region studied
here, i.e. $E_{cm}$ = 300 to 370 keV. This function has been folded with the
obtained target profiles and convoluted with a Gaussian function in order to
include the detector resolution. The cross section behavior should be the same in
all resulting spectra, hard and soft sum at the three energies $E_p$ = 360, 380
and 400 keV. Therefore, all spectra were fitted simultaneously using the three
polynomial coefficients as free parameters. The background was assumed to behave
linearly and the Gaussian peak at the resonance energy was also fitted to obtain
the widths for the Gaussian convolution. The overall reduced $\chi ^2$ was 1.15.

\begin{figure}[ptb]
    \centering
        \includegraphics[angle=-90,width=0.9\textwidth]{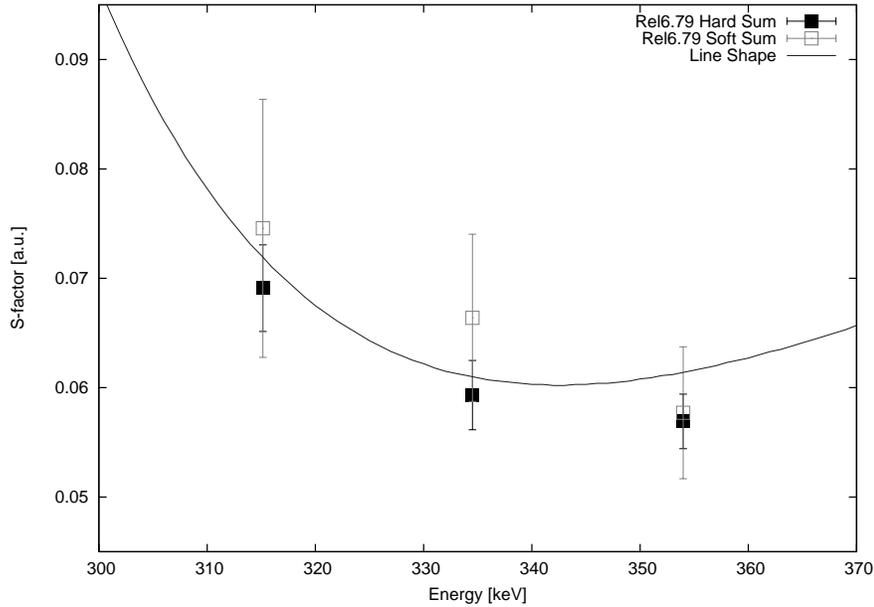}
    \caption{Preliminary results in arbitrary units for the relative analysis method (hard sum,
soft sum) and the line shape analysis (see text for details).}
    \label{fig:4}
\end{figure}

\paragraph{}
Figure \ref{fig:4} shows the astrophysical S--factor in arbitrary units, obtained
from different relative analysis. The results for both the hardware and software
sum relative to the 6.79 MeV transition are shown together with the results of the
shape analysis. The agreement is remarkable considering the fact that only the
efficiency is the common parameter based on the fitted stoichiometry. This shows
that the analysis methods described above are reliable.

\section*{Acknowledgments}
This work was supported by INFN and in part by the European Union (TARI
RII3-CT-2004-506222), the Hungarian Scientific Research Fund (K68801) and the
German Federal Ministry of Education and Research (05CL1PC1-1).

\end{document}